\documentclass[aps,prl,twocolumn,groupedaddress,showpacs]{revtex4}

\begin{document}

\title{Anomalous hydrodynamics and ``normal'' fluids in rapidly rotating BECs}

\author{A. Bourne$^{(1)}$, N. K. Wilkin$^{(1,2)}$ and J.M.F. Gunn$^{(1,2)}$}

\affiliation{$^{(1)}$School of Physics and Astronomy, University of
Birmingham, Edgbaston, Birmingham B15 2TT, U. K.}
\affiliation{$^{(2)}$ Laboratoire Physique Th\'eorique et Mod\`eles
Statistiques, B\^atiment 100, Universit\'e Paris-Sud, 91405 Orsay,
France}

\begin{abstract}
In rapidly rotating bose systems we show that there is a region of
anomalous hydrodynamics whilst the system is still condensed, which
coincides with the mean field quantum Hall regime. An immediate
consequence is the absence of a normal fluid in any conventional
sense. However, even the superfluid hydrodynamics is not described
by conventional Bernoulli and continuity equations. We show there
are kinematic constraints which connect spatial variations of
density and phase, that the positions of vortices are not the
simplest description of the dynamics of such a fluid (despite their
utility in describing the instantaneous state of the condensate) and
that the most compact description allows solution of some
illuminating examples of motion. We demonstrate, inter alia, a very
simple relation between vortices and surface waves. We show the
surface waves can form a ``normal fluid" which absorbs energy and
angular momentum from vortex motion in the trap. The time scale of
this process is sensitive to the initial configuration of the
vortices, which can lead to long-lived vortex patches - perhaps
related to those observed at JILA.
\end{abstract}


\maketitle

The area of rapidly rotating Bose Einstein condensates was one of
the first to produce predicted phenomena quite distinct to the
analogous condensed matter system, $^4$He. Initially an instability
was found\cite{WGS} to a non-condensed, Laughlin, state for
repulsive effective interactions between the atoms with sufficiently
high angular momentum, and the production\cite{WGS} of a fragmented
condensate in the rotating attractive case. Both of these novel
features occur in a regime where the atoms reside in restricted set
of single particle states, the ``Lowest Landau level" (LLL), defined
below. Subsequently it has been understood that when the atoms
reside in the Lowest Landau level at intermediate amounts of angular
momentum mean field theory provides a good description (Mean field
quantum Hall regime or MFQHR). In that regime the ground state is a
vortex lattice\cite{BR,Ho1} and it has been shown that other phases
occur en route to the Laughlin state in the correlated
domain\cite{CW,WG,CWG} as the amount of angular momentum in
increased. The nature of these transitions in the thermodynamic
limit\cite{Sinova} and in finite systems remains a very active area
of research, extending now into anisotropic traps\cite{Sinha,Matt}.

Following the pioneering studies\cite{Madison,Abo} of the vortex
lattice, current experiments are in\cite{Schweikhard04} or
near\cite{Dalibard04,Coddington} the mean field quantum Hall regime.
The correlated regime is still to be investigated. The experiments
find that the vortex lattice becomes very soft with an increasingly
lengthy period for recovery after disruption. In this Letter we show
the underlying hydrodynamics in the LLL is very unconventional and
will be a contributory factor to understanding these experimental
observations. There has been considerable theoretical
discussion\cite{Sinova,Baym} about the conventional or
unconventional nature of excitations in the vortex lattice; this
Letter provides a formulation of the hydrodynamics underlying such
excitations.

The defining characteristic of the wave functions, $\psi (x,y)$,
belonging to the lowest Landau level is that $\psi (z,z^*)= f(z){\rm
e}^{-|z|^2/2}$, where $z=x+{\rm i}y$ and $f(z)$ is analytic (or
entire) in $z$. $\psi(z,z^*)$ describes the system in the plane
perpendicular to the axis of rotation. In general $f(z,t)$ is
time-dependent. The standard basis set, reflecting the analyticity
of $f(z)$, is $\{(z^m/\sqrt{\pi m!})\thinspace {\rm
e}^{-|z|^2/2}\}$, where $m=0,1,\cdots$.

To derive hydrodynamic equations from the Gross-Pitaevskii equation
one conventionally represents the wave-function,
$\psi=\sqrt{\rho}\thinspace {\rm e}^{{\rm i}\phi}$, i.e. in terms of
its modulus, $\sqrt{\rho}$, and argument $\phi$. $\rho$ is
interpreted as the density and $\phi$ is the order parameter, and in
the nonrotating case is the velocity potential of the fluid. In the
case of LLL hydrodynamics this motivates a representation of
$f(z,t)$ as $f(z,t)=\sqrt{{\tilde \rho}}\thinspace {\rm e}^{{\rm
i}\phi}$, since the Gaussian factor is time-independent. The
analyticity of $f$ (i.e. $\partial f/\partial z^*=0$) leads to
``polar" analogues of the Cauchy-Riemann conditions for the
derivatives of ${\tilde \rho}$ and $\phi$:
\begin{eqnarray*}{\textstyle \frac{1}{2}}\frac{\partial
{\tilde \rho}}{\partial x} &=&{\tilde \rho} \thinspace
\frac{\partial \phi}{\partial y} \hbox{ and } {\textstyle
\frac{1}{2}}\frac{\partial {\tilde \rho}}{\partial y} =- {\tilde
\rho} \thinspace \frac{\partial \phi}{\partial
x}\\
\Leftrightarrow\thinspace \frac{\partial \ln \sqrt{{\tilde
\rho}}}{\partial x}&=& \frac{\partial \phi}{\partial y} \hbox{ and }
\frac{\partial \ln\sqrt{{\tilde \rho}}}{\partial y} =-
\frac{\partial \phi}{\partial x}\end{eqnarray*}

The importance of these relations is that they constrain
kinematically the spatial variation of the density (via $\rho =
{\tilde \rho}\thinspace {\rm e}^{-|z|^2}$) and phase unlike
conventional superfluid hydrodynamics. In addition, non-locality due
to projection of the interaction term to the LLL affects both the
continuity and ``Bernoulli" equations, making them difficult to use.

With these shortcomings in mind, we incorporate the analyticity of
the wave-function from the start. There are clearly two
complementary representations\cite{Tesan,BR,Ho1} which accomplish
this. Firstly the Taylor expansion of $f(z,t)$ may be written as:
$$\psi(z,z^*, t) = f(z,t){\rm e}^{-|z|^2/2} = \sum_{m=0}^\infty a_m(t) \thinspace z^m {\rm
e}^{-|z|^2/2}.$$ By the fundamental theorem of algebra, we may
factorise the polynomial in terms of its roots $\zeta_\alpha (t)$
(\onlinecite{Ho1}):
$$\psi(z,z^*, t) =  S^{-1/2} \prod_{\alpha = 1}^\infty
(\zeta_\alpha(t) - z)\thinspace {\rm e}^{-|z|^2/2}$$ where $S$ is
the normalisation integral. The points $\zeta_\alpha (t)= X_\alpha
(t) +{\rm i}Y_\alpha(t) $ are the vortex positions, or coordinates,
at time $t$.

We immediately see that there is an important distinction between a
condensate wave-function residing in the LLL and a general
condensate wave-function. In the former case the wave-function is {\it
completely specified} by the position of the vortices. In
particular, the density is specified by the vortex positions - there
are no independent sound waves (long wavelength Bogoliubov
quasiparticles), unlike the general case. There is thus {\it no
possibility} of the existence of a normal fluid component in a
conventional sense.

Since the vortices uniquely determine the wave-function in the LLL,
it is natural to re-express the Hamiltonian in terms of the vortex
coordinates,  $\zeta_\alpha (t)$. We use the standard model
Hamiltonian for $N$ particles in a rotating reference frame:
$${\cal H}= -{\textstyle \frac{1}{2}}\sum_{n=1}^{N}\nabla^2_n +
{\textstyle \frac{1}{2}}\sum_{n=1}^{N}r^2_n + {\textstyle
\frac{1}{2}}\eta \sum_{n\ne n'=1}^{N}\delta({\bf r}_n-{\bf r}_{n'})
- \omega \sum_{n=1}^{N}L^z_n$$ Units of length, $\ell_0$, and
energy, $\hbar \omega_0$, are those provided by the harmonic trap;
angular momenta, $L^z_n$ are scaled by $\hbar$. There are two
remaining dimensionless parameters. Firstly, $\omega$, is the
angular velocity of the rotating frame divided by the natural
frequency of the harmonic trap. Secondly the coupling constant,
$\eta = 4\pi {\bar n}a\ell_0^2$.

Then the Hamiltonian takes the form (assuming $M$ vortices):
${\cal
V}$, inter-particle potential, ${\cal U}$ energies and the rotating
frame term, ${\cal R}$, are defined as (using ${\dot {}}={\rm
d}/{\rm d}t$):
$$ S= \pi N \sum_{m}^M |P_{M-m}^*(\zeta)|^2 m!$$
$${\cal K}={\cal V} = \frac{1}{2}\pi N \sum_{m}^M
|P_{M-m}^*(\zeta)|^2 (m+1)!$$
$${\cal R} = \pi N \omega \sum_{m}^M |P_{M-m}^*(\zeta)|^2 m
m!$$ \begin{eqnarray*}{\cal U}= &\frac{\lambda}{4} N(N-1) S^{-2}
\sum_{m,n,p,q=0}^M
P_{M-m}^*(\zeta)P_{M-n}^*(\zeta)\\
&P_{M-p}(\zeta)P_{M-q}(\zeta) (p+q)!2^{-(p+q)}\delta_{m+n,p+q}
\end{eqnarray*}where the quantity $P_{n}(\zeta)= \sum_{i_1<i_2<\cdots
<i_n} \zeta_{i_1}\zeta_{i_2}\cdots \zeta_{i_n}$ is the $n$th
symmetric polynomial in the variables $\zeta_\alpha$.

It is instructive to contrast this expression with the corresponding
expression for vortices in an incompressible medium in a container
of size $R$:
$${\cal H}= - {\textstyle \frac{1}{2}} \Gamma^2 n_0\sum_{\alpha\ne\beta =1}^M\ln
|\zeta_\alpha - \zeta_\beta| -\omega \Gamma n_0
(R^2-|\zeta_\alpha|^2) + \thinspace \mbox{images}$$ where $\omega$
is the unscaled angular velocity of the rotating frame, $\Gamma$ is
the circulation of the vortices (assumed to be identical) and $n_0$
is the constant density.

There are two striking differences. Firstly \cite{Tesan}, the
interaction between the vortices is analytic and is not pairwise,
with multi (up to $M$-) vortex contributions. Secondly the
expression for the angular momentum of the fluid in terms of the
vortices (entering the term involving $\omega$) is not a sum of
single vortex contributions but is collective in nature in the LLL
case.

To determine the equations of motion of the condensate constrained
to reside in the LLL, we employ a variational
formulation\cite{Kramer,Fetter} where the LLL wavefunction is a
trial function leading to the Lagrangian, ${\cal L}= {\cal T} -
{\cal H}$, where (assuming $\psi(z,z^*,t)$ is normalised)
$${\cal T}= \pi N \sum_{m=0}^M m! \frac{\rm i}{2}
(P_{M-m}^*(\zeta) {\dot P}_{M-m}(\zeta) - P_{M-m}(\zeta) {\dot
P}_{M-m}^*(\zeta))$$ If the corresponding action is varied with
respect to the $\zeta_\alpha$ the resulting equations of motion for
the vortex coordinates are rather complicated. However the form of
$\cal L$ shows that the vortex coordinates always enter in the form
of the symmetric polynomials, $P_n(\{\zeta_\alpha\})$. This suggests
that their numerical values should be used as the primary variables.
Since, up to a sign, the $P_n(\{\zeta_\alpha\})$ are the
coefficients in the polynomial whose roots are the $\zeta_\alpha$,
there is a one-to-one relation between the vortex coordinates and
the symmetric polynomial values. The dynamics are most simply
expressed in the variables $\rho_m (t)$ and $\theta_m(t)$ implicitly
defined by:
$$(-1)^m P_{M-m}(\zeta)=a_m =\sqrt{\frac{\rho_m}{\pi m!}} \thinspace {\rm e}^{{\rm
i}\theta_m}$$ In particular normalisation can be easily enforced via
the $\rho_m$.

In terms of $\rho_m$ and $\theta_m$ we find (\thinspace $\dot{}$
indicates a time derivative and $\bar{\lambda} =
\lambda(N-1)/4\pi$):
\begin{eqnarray*}{\cal L}&=& N \left\{ S^{-1}\sum_{m=0}^M \rho_m
[-\dot{\theta}_m - (1+m[1-\omega])] \right. \\
&-&\left.\bar{\lambda}\sum_{m,n,p,q=1}^M \sqrt{\rho_m \rho_n \rho_p
\rho_q} {\rm e}^{{\rm
i}[(\theta_m+\theta_n)-(\theta_p+\theta_q)]}\right.\\
&\phantom{-}&\left.
\sqrt{\frac{(m+n)!}{2^{m+n}m!n!}}\sqrt{\frac{(p+q)!}{2^{p+q}p!q!}}
\delta_{p+q,m+n}\right\}\end{eqnarray*} where
$S=\sum_{m=0}^M\rho_m$.

Differences between incompressible\cite{Saff} or
Thomas-Fermi\cite{Fetter2} and LLL vortex dynamics are already
apparent in the case of $m$ vortices symmetrically arranged and
equidistant with respect to the centre of the trap. Up to a phase,
the vortices are in the directions of the $m$th roots of unity in
the $\zeta$-plane, and hence the wave-function is of the form $\phi
(z) \sim (a_0(t) + a_m(t)z^m)$. We will initially restrict ourselves
to a reduced Lagrangian only involving the $\rho_m$, $\theta_0$ and
$\theta_m$ variables (as $\rho_0$ can be eliminated by the
normalisation constraint $1=\rho_0+\rho_m$):
\begin{eqnarray*}\frac{\cal L}{N} &=& {-\dot \theta}_0 -1-\bar{\lambda}\\
 &+& \rho_m\left([-{\dot \theta}_m - {\dot \theta}_0]-m[1-\omega]
-\bar{\lambda}[2^{-(m-2)}-2]\right)\\&& +\rho_m^2 \bar{\lambda}
\left(2^{-(m-2)}-1- \frac{(2m)!}{(m!)^2}
2^{-2m}\right)\end{eqnarray*} $\theta_0$ is not determined and its
time derivative may be interpreted as the zero of energy in the
original quantum problem.

The significant Euler-Lagrange equation is that associated with
$\rho_m$ (the equation for $\theta_m$ expresses the unitarity of the
original time evolution, in this case ${\dot \rho_m}=0$), yielding
the excitation frequency $\omega_m$:
\begin{eqnarray*}\omega_m&=&{-\dot \theta}_m + {\dot \theta}_0\\
&=& m(1-\omega) - 2\bar{\lambda}(1-2^{-(m-1)})\\
 &&+
\rho_m 2\bar{\lambda}\left(1+\frac{(2m)!}{(m!)^2}
2^{-2m}-2^{-(m-2)}\right)\end{eqnarray*} This expression describes,
in a unified manner, several physical phenomena: surface wave
excitations, instabilities and anomalous vortex dynamics.

We may re-express the equation for $\omega_m$ in terms of the radial
distance of the vortices from the centre, $\zeta$, as $\rho_m=
1/(1+\zeta^{2m}/m!)$. This implies a natural scale for the radial
vortex positions of $\zeta_m = (m!)^{1/(2m)}\sim \sqrt{m}$, which
accords with the extent of the condensate with $m$ quanta of angular
momentum. For $\zeta \lesssim \zeta_m$, the vortices are ``inside"
the condensate, and for $\zeta\gtrsim \zeta_m$ the vortices are in
the evanescent tail.

Thus in the limit that the vortices are very far from the centre of
the trap (i.e. $\rho_m\to 0$) we recover the linear surface
excitation frequencies derived by Kavoulakis et al \cite{Kav}. We
are thus immediately provided with a representation of surface waves
in the LLL of angular momentum $m$ in terms of $m$ vortices. A
relationship between vortices and surface waves in the Thomas-Fermi
limit has been discussed by Anglin \cite{Anglin} and Tsubota et
al\cite{Tsubota}.

To obtain the equilibrium radial distance of the vortices,
$\zeta_{\rm eq}(\omega)$, we set $\omega_m=0$ so that the vortices
are stationary in the rotating frame. We find $\zeta_{\rm
eq}(\omega)= \infty$ for $\omega$ less than a critical rotation
frequency, $\omega_{\rm c}^{(m)}=1-2\bar{\lambda}
\frac{1-2^{-(m-1)}}{m}$. For $\omega>\omega_{\rm c}^{(m)}$, we see:
$$\frac{\zeta_{\rm eq}^{2m}(\omega)}{m!}=
\frac{2{\bar\lambda}}{ m(\omega-\omega_{\rm
c}^{(m)})}\left(1+\frac{(2m)!}{(m!)^2}2^{-2m}-2^{-(m-2)}\right)$$
These critical frequencies do not have a simple physical
interpretation as shape instabilities of the condensate - as, for
example, $\omega_{\rm c}^{(2)}$ coincides with the frequency of a
first order transition of a single vortex to the centre of the trap.
This is unlike the Thomas-Fermi limit\cite{Sinha2}.

One might expect the dynamics of the vortices when they are
``inside" the trap, $\zeta \lesssim \zeta_m$, to be reminiscent to
incompressible vortex dynamics as the phases of the wave-functions
are identical. However the LLL dynamics is completely different. For
example as the separation of the vortices vanishes, the frequency
$\omega_m(\zeta=0)$ obeys:
\begin{eqnarray*}\omega_m(\zeta=0)&=&m(1-\omega)+2\bar{\lambda}2^{-2m}\left(\frac{(2m)!}{(m!)^2}-2\right)\\
&\raisebox{-1ex}{$ \stackrel{\sim}{\scriptstyle
m\to\infty}$}&m(1-\omega)+\frac{2\bar{\lambda}}{ \sqrt{\pi m}}\\
\end{eqnarray*}
It does not diverge, in stark contrast with the incompressible case.
There the frequency, $\Omega$, of rotation of such a polygonal array
of $m$ vortices, each with circulation $\Gamma$, a distance $\zeta$
from the centre is\cite{Saff} (in the laboratory frame) $\Omega
=\Gamma(m-1)/(4\pi \zeta^2)$.

We will now show that energy and angular momentum can be deposited
in surface waves (vortices at large distances) by vortices near the
centre of the trap. In other words the surface waves can in part
play the role of a ``normal fluid". Consider $n$ vortices which are
all initially very near the centre of the trap. A linearised
approximation (where $\rho_n \simeq 1$) allows coupling to all the
angular momentum channels which conserve angular momentum (i.e.
$m\le 2n$). This implies we will have $n$ additional vortices, which
will turn out to be at large distances from the centre of the trap.
Because of angular momentum conservation $\rho_{n-m}-\rho_{n+m}$ is
a constant, which we choose to be zero. Then the equations for
$R_m=\frac{1}{2} (\rho_{n-m}+\rho_{n+m})$ and
$\sigma_m=(\theta_{n-m}-\theta_n)+(\theta_{n+m}-\theta_n)$ are:
\begin{eqnarray*}
{\dot R}_m&=& -\Lambda_m \thinspace R_m \sin \sigma_m\\
{\dot \sigma}_m &=& \chi_m -\Lambda_m \cos \sigma_m
\end{eqnarray*}
Where
\begin{eqnarray*}
\Lambda_m&=& 4\bar{\lambda}\frac{(2n)!}{2^{2n} n!\sqrt{(n+m)!(n-m)!}}\\
\chi_m &=& 4\bar{\lambda}\frac{1}{2^{2n}n!}\left[\frac{(2n)!}{n!} -
\frac{(2n-m)!}{2^{-m}(n-m)!} -\frac{(2n+m)!}{2^{+m}(n+m)!}\right]
\end{eqnarray*}
These equations are readily integrated and one finds exponentially
unstable growth for $n+c \sqrt{n} \gtrsim m > n+1$ and $n-1
>m \gtrsim n-c\sqrt{n}$, in the limit where $n\gg 1$, where
$c=2(-\ln[\sqrt{2}-1])^{1/2}$. The extent of this unstable region is
determined by the spatial overlap of the states with the state $n$.
Oscillatory behaviour occurs for other values of $m$. The time
constant, $\tau_m$, for the growth is
$\tau_m^{-1}=\frac{1}{2}\sqrt{\Lambda_m^2-\chi_m^2}$. For the most
divergent cases, $m*\simeq n\pm
2(\ln[2/(\sqrt{5}-1)])^{1/2}\sqrt{n}$, we find
$\tau^{-1}_{m*}\raisebox{-1ex}{$ \stackrel{\propto}{\scriptstyle
n\to\infty}$} \bar{\lambda}/\sqrt{n}$.

If we assume that initial values of $\rho_m$, $\rho_m(0)$, are of
the order of $\epsilon \ll 1$, and then the unstable modes grow by
time $t$ to $\rho_{m^*}\sim \delta$ where $\epsilon \ll \delta \ll
1$, then we can deduce the corresponding vortex positions from the
crude representation of the polynomial:
\begin{eqnarray*}z^n &+& \delta(z^{n-\sqrt{n}} +
\cdots + z^{n-1} + z^{n+1}+\cdots +z^{n+\sqrt{n}})\\
&+&\epsilon (1+z+\cdot\cdot +z^{n-\sqrt{n}-1}+
z^{n+\sqrt{n}+1}+\cdot\cdot +z^{2n})=0\end{eqnarray*} Noting that if
$z$ is a solution so is $1/z$, let us look for solutions for large
$|z|$. Keeping the largest terms multiplying each of the two small
parameters, we obtain
$$1+x+\frac{\epsilon}{\delta^{\sqrt{n}}}x^{\sqrt{n}}\simeq 0$$
where $z^{\sqrt{n}}=x/\delta$. Balancing the middle term with either
the final term or with unity, we find $\sqrt{n}$ approximate
solutions $z^{\sqrt{n}}=-1/\delta$ and $n-\sqrt{n}$ approximate
solutions $z^{\sqrt{n}(\sqrt{n}-1)}=-\delta/\epsilon$. (We assume
$\epsilon\ll \delta^{\sqrt{n}}$.)

Geometrically, the roots form four concentric rings, two inside the
trap and two outside. The closest and the most remote rings do not
change their moduli significantly, but the rings with moduli
$|\delta|^{\pm 1/\sqrt{n}}$ change with time. The inner ring of
$\sqrt{n}$ vortices, moving outwards and the outer ring (with the
same number) moving inwards. Each member of the inner ring has a
corresponding member of the outer ring with the same argument. This
demonstrates the exchange of angular momentum  between the inner and
outer rings.

This description assumes that one draws the initial conditions from
the same probability distribution for each coefficient in the
polynomial - which leads to rings of roots\cite{Bog,Forr}. However
for initial vortex positions forming a cloud distributed normally
around the origin in the complex plane, then the coefficients of the
polynomial are very strongly peaked\cite{Bog} at $m\simeq n/2$. Such
a configuration will evolve much more slowly as it has negligible
initial weight in the unstable modes, $n-\sqrt{n}\lesssim m<n$. A
simple estimate using the expressions above yields time scales for
evolution $t_{\rm ring}\sim \sqrt{n}\ln n$ and $t_{\rm cloud}\sim
n^{3/2}$. It is tempting to compare this distinction with the JILA
experiments\cite{Engels} where long-lived concentrations of vortices
were observed. Although the experiments were not conducted under LLL
conditions on average, in the low-density environment of the
vicinity of the vortices the dynamics might be of a LLL nature.

NKW and JMFG would like to thank KITP, University of California at
Santa Barbara, the Aspen Center for Physics and Gora Shlyapnikov for
hospitality at Universit\'e Paris-Sud while parts of this work was
performed and thank him, Nigel Cooper and Martin Long for several
helpful discussions. This work was also supported by EPSRC grants
GR/R80865 (NKW) and GR/R00920 (JMFG).


\begin{thebibliography}{10}
\bibitem{WGS} N.K. Wilkin, J.M.F. Gunn and R.A. Smith, Phys. Rev.
Lett. {80}, 2265 (1998).
\bibitem{Tesan} Z.
Tesanovic and L. Xing, Phys. Rev. Lett. {\bf 67}, 2729 (1991); Z.
Tesanovic, Phys. Rev. B {\bf 44} 12635 (1991).
\bibitem{BR}D.A. Butts and D.S. Rokhsar, Nature (London),
{\bf 397}, 297 (1999).
\bibitem{Ho1}T.-L. Ho. Phys. Rev. Lett. {\bf 87}, 060403 (2001)
\bibitem{CW} N.R. Cooper and N.K. Wilkin, Phys. Rev. B 60,
R16279, (1999).
\bibitem{WG} N.K. Wilkin and J.M.F. Gunn Phys. Rev. Lett. {\bf 84},
6 (2000).
\bibitem{CWG} N.R. Cooper, N.K. Wilkin and J.M.F. Gunn, Phys. Rev.
Lett. {\bf 87}, 120405 (2001).
\bibitem{Sinova} J. Sinova, C.B. Hanna and A.H. Macdonald, Phys. Rev.
Lett. {\bf 89}, 030403 (2002).
\bibitem{Sinha} S. Sinha and G.V. Shlyapnikov, Phys. Rev. Lett. {\bf 94},
150401 (2005)
\bibitem{Matt} M.I. Parke and N.K. Wilkin, (To be published).
\bibitem{Madison} K. W. Madison, F. Chevy, W. Wohlleben and J.
Dalibard, Phys. Rev. Lett. {\bf 84,} 806 (2000).
\bibitem{Abo} J.R. Abo-Shaeer, C. Raman, J.M. Vogels and W.
Ketterle, Science {\bf 292}, 476 (2001).
\bibitem{Schweikhard04}V. Schweikhard, I. Coddington. P. Engels,
V.P. Mogendorff and E.A. Cornell, Phys. Rev. Lett. {\bf 92}, 040404
(2004).
\bibitem{Dalibard04}V. Bretin, S. Stock, Y. Seurin and J. Dalibard,
Phys. Rev. Lett. {\bf 92}, 050403 (2004).
\bibitem{Coddington}I. Coddington, P. Engels, V. Schweikhard and
E.A. Cornell, Pys. Rev. Lett. {\bf 91}, 100402 (2003).
\bibitem{Baym} G. Baym, Phys. Rev. Lett. {\bf 91}, 110402 (2003);
G. Baym, Phys. Rev. A {\bf 69}, 043618 (2004).
\bibitem{Kramer} P. Kramer and M. Saraceno {\em Geometry of the
Time-Dependent Variational Principle in Quantum Mechanics}, Lecture
Notes in Physics {\bf 140} (Springer-Verlag, Berlin, 1981).
\bibitem{Fetter}M. Linn and A.L. Fetter, Phys. Rev. A {\bf 61}, 063603 (2000).
\bibitem{Kav} G. M. Kavoulakis, B. Mottelson and C. J. Pethick, Phys. Rev. A {\bf 62},
063605 (2000).
\bibitem{Anglin} J.R. Anglin, Phys. Rev. Lett. {\bf 87}, 240401
(2001).
\bibitem{Tsubota}M. Tsubota, K. Kasamatsu and M. Ueda, Phys. Rev. A {\bf 65}, 023603
(2002); K. Kasamatsu, M. Tsubota and M. Ueda, Phys. Rev. A {\bf 67},
033610 (2003).
\bibitem{Saff}P.~G. Saffman, {\em Vortex dynamics} (Cambridge University Press, Cambridge, 1995), p. 119.
\bibitem{Fetter2} J-K Kim and A.L. Fetter, Phys. Rev. A {\bf 70}, 043624
(2004).
\bibitem{Sinha2}S. Sinha and Y. Castin, Phys. Rev. Lett. {\bf 87},
190402 (2001).
\bibitem{Bog}E.B. Bogomolny, O. Bohigas, P. Leboeuf, Phys. Rev.
Lett. {\bf 68}, 2726 (1992); E.B. Bogomolny, O. Bohigas, P. Leboeuf,
J. Stat. Phys. {\bf 85}, 639 (1996).
\bibitem{Forr}P.J. Forrester, G. Honner, J. Phys. A {\bf 32} 2961
(1999).
\bibitem{Engels}P. Engels, I. Coddington, P.C. Haljan, V. Schweikhard, E.A.
Cornell, Phys. Rev. Lett. {\bf 90 } 170405 (2003).

\end{thebibliography}
\end{document}